\documentclass[120pt, final, journal, letterpaper, twocolumn]{IEEEtran}
\usepackage{amsmath}
\usepackage{amssymb}
\usepackage{amsfonts}
\usepackage{caption}
\usepackage{graphicx}
\usepackage{epsfig}
\usepackage{epsf}
\usepackage{subfigure}
\usepackage{psfrag}
\usepackage{cite}
\usepackage{latexsym}
\usepackage{url}
\usepackage{color}
\usepackage{bm}
\usepackage{calc}
\usepackage{cases}
\usepackage{empheq}
\usepackage{algorithm}
\usepackage{subcaption}
\usepackage{algpseudocode}
\usepackage{graphics}
\begin{document}
\captionsetup[figure]{name={Fig.},labelsep=period}
\title{Sensing-Assisted Channel Prediction in Complex Wireless Environments: An LLM-Based Approach}
\author{\IEEEauthorblockN{Junjie He, Zixiang Ren, Jianping Yao, Han Hu, Tony Xiao Han, and Jie Xu}
\thanks{
J. He, Z. Ren, and J. Xu are with the
School of Science and Engineering (SSE), the Shenzhen Future Network of Intelligence Institute (FNii-Shenzhen), and the Guangdong Provincial Key Laboratory of Future Networks of Intelligence, The Chinese University of Hong Kong, Shenzhen, Guangdong 518172, China (E-mails: 2112303110@mail2.gdut.edu.cn, rzx66@mail.ustc.edu.cn, xujie@cuhk.edu.cn).


J. Yao is with School of Information Engineering, Guangdong University of Technology, Guangzhou 510006, China(e-mail: yaojp@gdut.edu.cn).

H. Hu is with the School of Information and Electronics, Beijing Institute of Technology, Beijing 100081, China (e-mail: hhu@bit.edu.cn).

T. X. Han is with Huawei Technologies Company Ltd., Shenzhen 518129, China (e-mail: tony.hanxiao@huawei.com).

J. Xu is the corresponding author.}
}
\vspace{-1cm}
\maketitle
\begin{abstract}
This letter studies the sensing-assisted channel prediction for a multi-antenna orthogonal frequency division multiplexing (OFDM) system operating in realistic and complex wireless environments. In this system, an integrated sensing and communication (ISAC) transmitter leverages the mono-static sensing capability to facilitate the prediction of its bi-static communication channel, by exploiting the fact that the sensing and communication channels share the same physical environment involving shared scatterers. Specifically, we propose a novel large language model (LLM)-based channel prediction approach, which adapts pre-trained text-based LLM to handle the complex-matrix-form channel state information (CSI) data. This approach utilizes the LLM’s strong ability to capture the intricate spatiotemporal relationships between the multi-path sensing and communication channels, and thus efficiently predicts upcoming communication CSI based on historical communication and sensing CSI data. Experimental results show that the proposed LLM-based approach significantly outperforms conventional deep learning-based methods and the benchmark scheme without sensing assistance.
\end{abstract}
\vspace{-10pt}
\section{Introduction}
Acquiring accurate channel state information (CSI) is becoming increasingly important for beamforming and precoding in multi-antenna wireless communication systems, especially when the antenna size becomes extremely large for future 6G networks \cite{cui.shuguang_large-scale_mimo}. Towards this end, channel prediction has drawn significant attention, which can leverage the temporal correlation between past and future channel states to forecast upcoming communication CSI based on historical observations \cite{4.yin2020addressing,liu2024llm4cp,transformer, convlstm}. In the literature, various channel prediction methods based on statistical (e.g., \cite{4.yin2020addressing}) and learning-based approaches (e.g., \cite{transformer,convlstm,liu2024llm4cp}) have been proposed, which rely on historical communication CSI as the sole input for prediction. 
With recent advances in integrated sensing and communication (ISAC) \cite{7.liu2022integrated}, leveraging the sensing capabilities of wireless transmitters has emerged as a promising solution to improve channel acquisition. In practice, mono-static sensing channels and bi-static communication channels share the same physical environment and exhibit overlapping multi-path components involving shared scatterers. As a result, the sensing CSI data also contain useful information about the behavior of communication channels, which can thus be utilized to facilitate the acquisition of communication CSI. There have been several prior works investigating sensing-assisted channel estimation \cite{ren2024sensing} and predictive beamforming \cite{8_liu_ywj, sensing-assisted_bf} to show the feasibility of this idea. However, these prior works \cite{ren2024sensing, 8_liu_ywj,sensing-assisted_bf} focused on simplified scenarios with only line-of-sight (LoS) channel paths or with limited non-LoS (NLoS) paths.

Different from prior works, this letter investigates the sensing-assisted channel prediction for a multi-antenna orthogonal frequency division multiplexing (OFDM) system operating in realistic and complex wireless environments, in which an ISAC transmitter utilizes both historical communication and sensing CSI to predict upcoming communication CSI. This, however, is a challenging task, as accurately modeling the underlying relationship between practical sensing and communication channels is inherently difficult. This difficulty stems from the dynamic nature of wireless environments, which are characterized by rich multi-path propagation, numerous scatterers, reflections, and continuous variations. 

To address this challenge in such complex wireless environments, we propose a novel large language model (LLM)-based sensing-assisted channel prediction approach. Pre-trained on vast datasets, LLMs excel at capturing intricate patterns within sequential data \cite{15.su2024large, timellm}. This makes them perfectly suitable for modeling the complex spatiotemporal dependencies between sensing and communication CSI. Motivated by this observation, this work presents the first attempt to apply LLM to sensing-assisted channel prediction. Specifically, we design a set of specialized modules for preprocessing, feature extraction and fusion, and output generation, enabling the adaptation of pre-trained text-based LLMs to handle the complex-matrix-form CSI data. In the fusion module, we utilize the  cross-attention to effectively fuse sensing and communication CSI, thereby enhancing the prediction process. Finally, experimental results show that the proposed LLM-based approach significantly outperforms conventional deep learning-based methods and the benchmark scheme without sensing assistance.



\section{System Model}
\begin{figure}
\setlength{\abovecaptionskip}{-0pt}
\setlength{\belowcaptionskip}{-20pt}
\centering
\includegraphics[width=9.0cm]{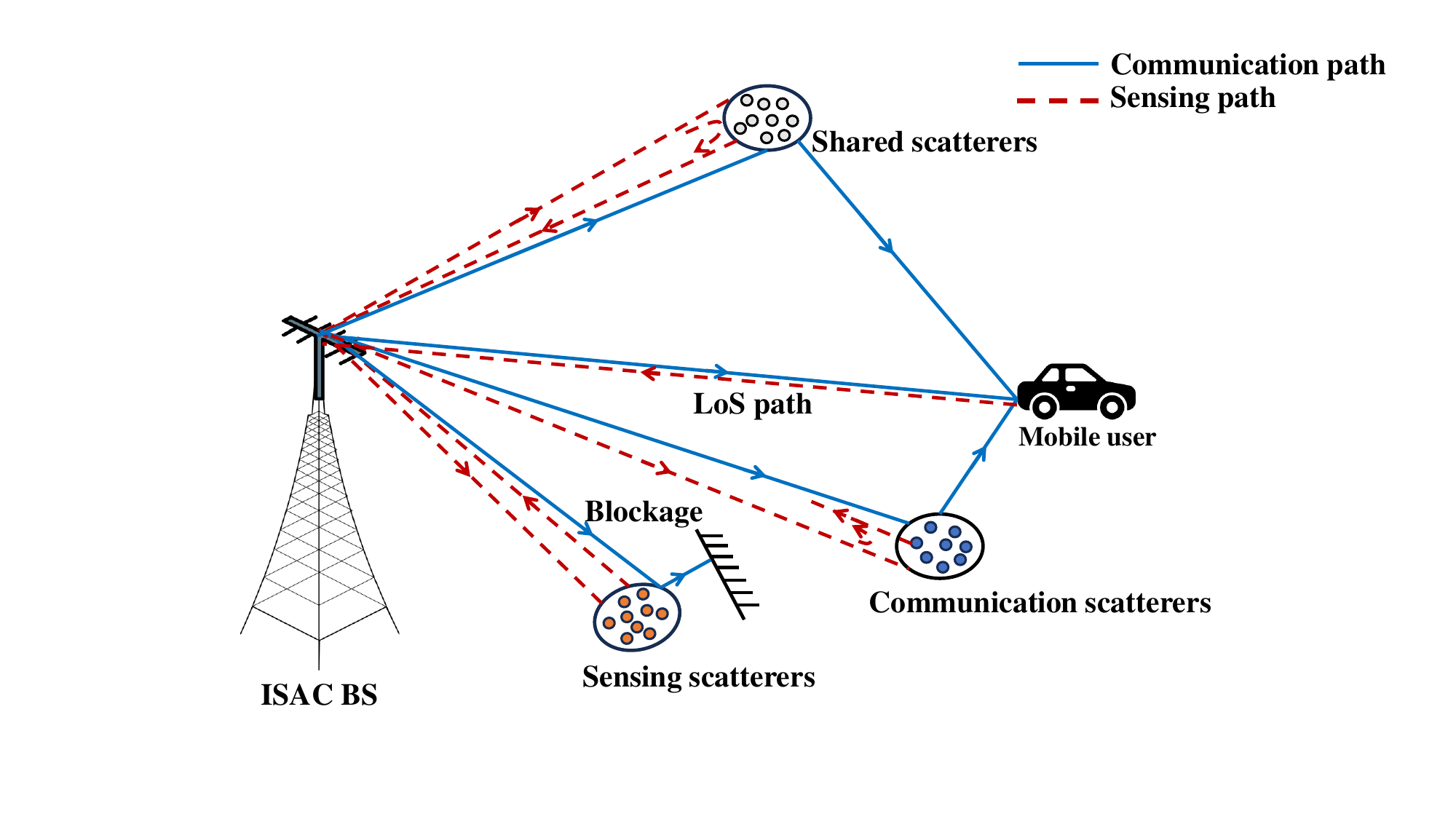}
\caption{Illustration of an example ISAC system in a street scene.}
\label{Fig.scene}
\end{figure}

We consider a basic multi-antenna OFDM ISAC system as shown in Fig. \ref{Fig.scene}, which includes a multi-antenna ISAC base station (BS) transmitter and a single-antenna mobile user (MU) receiver. The ISAC BS is equipped with a uniform planar array (UPA) of $N = N_{v}\times  N_{h} $ elements for sensing and communication, where $N_{v}$ and $N_{h} $ denote the numbers of vertical and horizontal antennas, respectively. In this system, the BS sends unified ISAC signals to communicate with the MU, and at the same time receives echo signals from surrounding environments for sensing. We consider quasi-stationary channel models, in which the communication and sensing channels remain unchanged at each slot but may change over different slots. We are interested in utilizing the sensing and communication CSI in the previous $P$ slots to predict the communication CSI in the upcoming $Q$ slots. 

As shown in Fig. \ref{Fig.scene}, the bi-static communication channel from the BS to the MU and the mono-static sensing channel from the BS to the environment and back to the BS are determined by environmental scatterers. Due to the shared physical propagation environment, scatterers can be classified into three types, i.e., shared, communication, and sensing scatterers. Shared scatterers contribute signal paths for both communication and sensing channels, whereas communication and sensing scatterers exclusively contribute to their respective channels. Let $\mathcal{N}_\mathsf{sha} = \{1, \ldots, N_0\}$, $\mathcal{N}_\mathsf{c} = \{N_0 + 1, \ldots, N_0 + N_1\}$, and $\mathcal{N}_\mathsf{s} = \{N_0 + N_1+ 1, \ldots, N_0 + N_1 + N_2\}$ denote the index sets of shared, communication, and sensing scatterers, with $N_0$, $N_1$, and $N_2$ being their corresponding numbers, respectively.\footnote{For ease of description, if the LoS signal path exists between the BS and the MU, then the MU itself can be viewed as a shared scatterer of both the communication and sensing channels.}



First, we consider the bi-static communication channel from the BS to the MU, which consists of the signal paths associated with both shared and communication scatterers. For each path $i$ related to scatterer $i\in \mathcal{N}_\mathsf{sha} \cup \mathcal{N}_\mathsf{c}$ at slot $t$, let $\alpha _{i}(t)$ denote the complex path coefficient, $\tau_{i}$ denote the propagation delay, $f_{\mathsf{d}, i}$ denote the Doppler frequency shift, and $\theta_{i}$ and $\varphi_{i}$ denote the elevation and azimuth angles of departure (AoDs), respectively. Accordingly, the steering vector associated with path $i$ is given by
\begin{align}\nonumber
\boldsymbol{a}(\theta_i, \varphi_i) = \boldsymbol{a}_{v}(\theta_i) \otimes \boldsymbol{a}_{h}(\theta_i, \varphi_i),
\end{align}
with $\boldsymbol{a}_{v}(\theta_i) = \frac{1}{N_{v}} \left[1, e^{j2\pi\frac{d_{v}}{\lambda}\sin\theta_i}, \ldots, e^{j2\pi\frac{d_{v}}{\lambda}(N_{v}-1)\sin\theta_i} \right]^{T}$ and $ \boldsymbol{a}_{h}(\theta_i, \varphi_i) = \frac{1}{N_{h}} \left[1, e^{j2\pi\frac{d_{h}}{\lambda}\cos\theta_i\sin\varphi_i}, \ldots,\right.$ $\left. e^{j2\pi\frac{d_{h}}{\lambda}(N_{h}-1)\cos\theta_i\sin\varphi_i}\right]^{T}$ 
denoting the steering vectors related to the elevation and azimuth AoAs, respectively, $\lambda$ denoting the wavelength,  and $d_v$ and $d_h$ denoting the spacing between adjacent vertical and horizontal antennas, respectively. Here, $j=\sqrt{-1}$, $\otimes$ denotes the Kronecker product, and superscript $T$ denotes the transpose. As such, the communication channel at slot $t$ and delay $\tau$ is given by
\begin{equation}
\begin{aligned}
        &\boldsymbol{h}_{\mathsf{c},t} (\tau )=\sum_{i\in \mathcal{N}_\mathsf{sha}} \alpha _{i}(t)e^{j2\pi f_{\mathsf{d},i }t }  \boldsymbol{a}(\theta _{i},\varphi _{i})\delta (\tau-\tau_{i}) \\
        &+\sum_{i\in \mathcal{N}_\mathsf{c}} \alpha _{i} (t)e^{j2\pi f_{\mathsf{d},i }t } \boldsymbol{a}(\theta _{i},\varphi _{i})\delta (\tau-\tau_{i}). 
\end{aligned}
\label{communication channel}
\end{equation}
Note that in (\ref{communication channel}), we assume that the complex coefficient $\{\alpha _{i}(t)e^{j2\pi f_{\mathsf{d},i }t}\}$ may change over different slots due to the movement of MU and scatterers, but other parameters $\{f_{\mathsf{d},i},\theta _{i},\varphi _{i}, \tau_{i}\}$ remain unchanged over the considered $P+Q$ slots due to their relatively slow changes. Therefore, the channel variations over slots are mainly due to the constructive and destructive combination of multiple random-phase signal paths, as commonly assumed in wireless communications \cite{goldsmith2005wireless}. Supposing that there are $K$ subcarriers for the OFDM system, we have the equivalent frequency-domain channel coefficient at each subcarrier $k\in\{1,\ldots,K\}$ as $\bar{\boldsymbol{h}}_{\mathsf{c},t}[k]$, which relates to $\{\boldsymbol{h}_{\mathsf{c},t}(\tau)\}$ based on discrete Fourier transform (DFT). 


Next, we consider the mono-static sensing channel at the BS. For each round-trip path $i$ related to scatterer $i\in \mathcal{N}_\mathsf{sha} \cup \mathcal{N}_\mathsf{s}$ at slot $t$, let $\beta_{i}(t)$, $\bar{\tau}_{i}$, $f_{\mathsf{d},i}$, $\theta_{i}$, and $\varphi_{i}$ denote the complex path coefficient, round-trip delay, Doppler frequency shift, and azimuth and elevation AoD/angles of arrival (AoAs), respectively. Accordingly, the sensing channel is given by
\setlength{\abovedisplayskip}{1pt}
\setlength{\belowcaptionskip}{1pt}
\begin{equation}
    \begin{aligned}
        &\boldsymbol{H}_{\mathsf{s},t} (\tau )=\sum_{i\in \mathcal{N}_\mathsf{sha}} \beta _{i} (t)e^{j2\pi f_{\mathsf{d},i}t } \boldsymbol{a}(\theta _{i},\varphi _{i}) \boldsymbol{a}^{T}(\theta _{i},\varphi _{i})\delta (\tau-\bar{\tau}_{i}) \\
        &+\sum_{i\in \mathcal{N}_\mathsf{s}} \beta _{i} (t)e^{j2\pi f_{\mathsf{d},i}t } \boldsymbol{a}(\theta _{i},\varphi _{i}) \boldsymbol{a}^{T}(\theta _{i},\varphi _{i})\delta (\tau-\bar{\tau}_{i}),
    \end{aligned}
\label{sensing channel}
\end{equation}
Based on $\{\boldsymbol{H}_{\mathsf{s},t}(\tau)\}$, we have the equivalent frequency-domain channel matrix at each subcarrier $k\in\{1,\ldots,K\}$ as $\bar{\boldsymbol{H}}_{\mathsf{s},t}[k]$.

\begin{figure}[t]
\setlength{\belowcaptionskip}{-5pt}
\centering
\begin{minipage}[t]{0.48\linewidth}
  \centering
  \includegraphics[width=\linewidth]{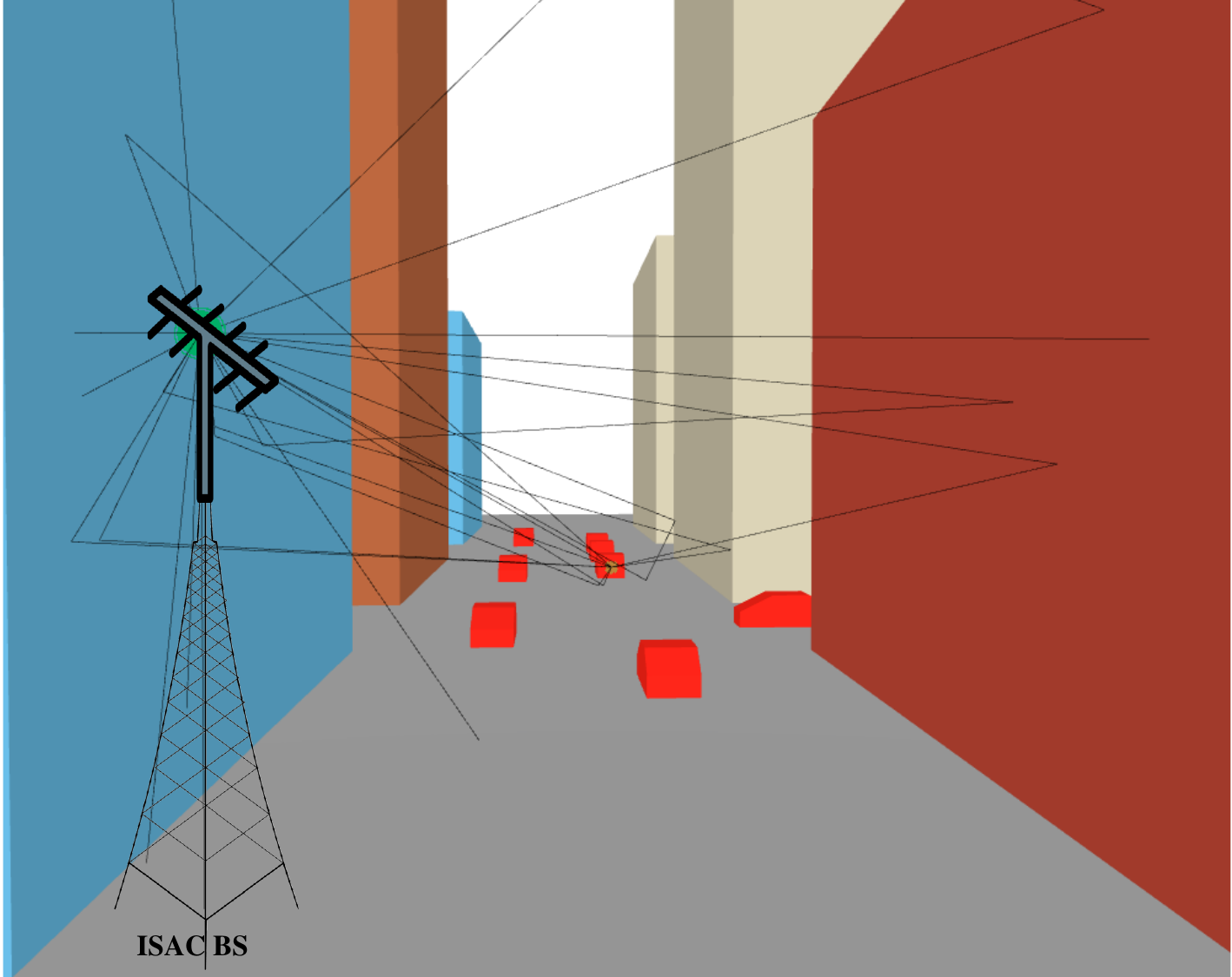}
  \caption*{(a) Simulation setup.}
\end{minipage}
\hfill
\begin{minipage}[t]{0.48\linewidth}
  \centering
  \includegraphics[width=\linewidth]{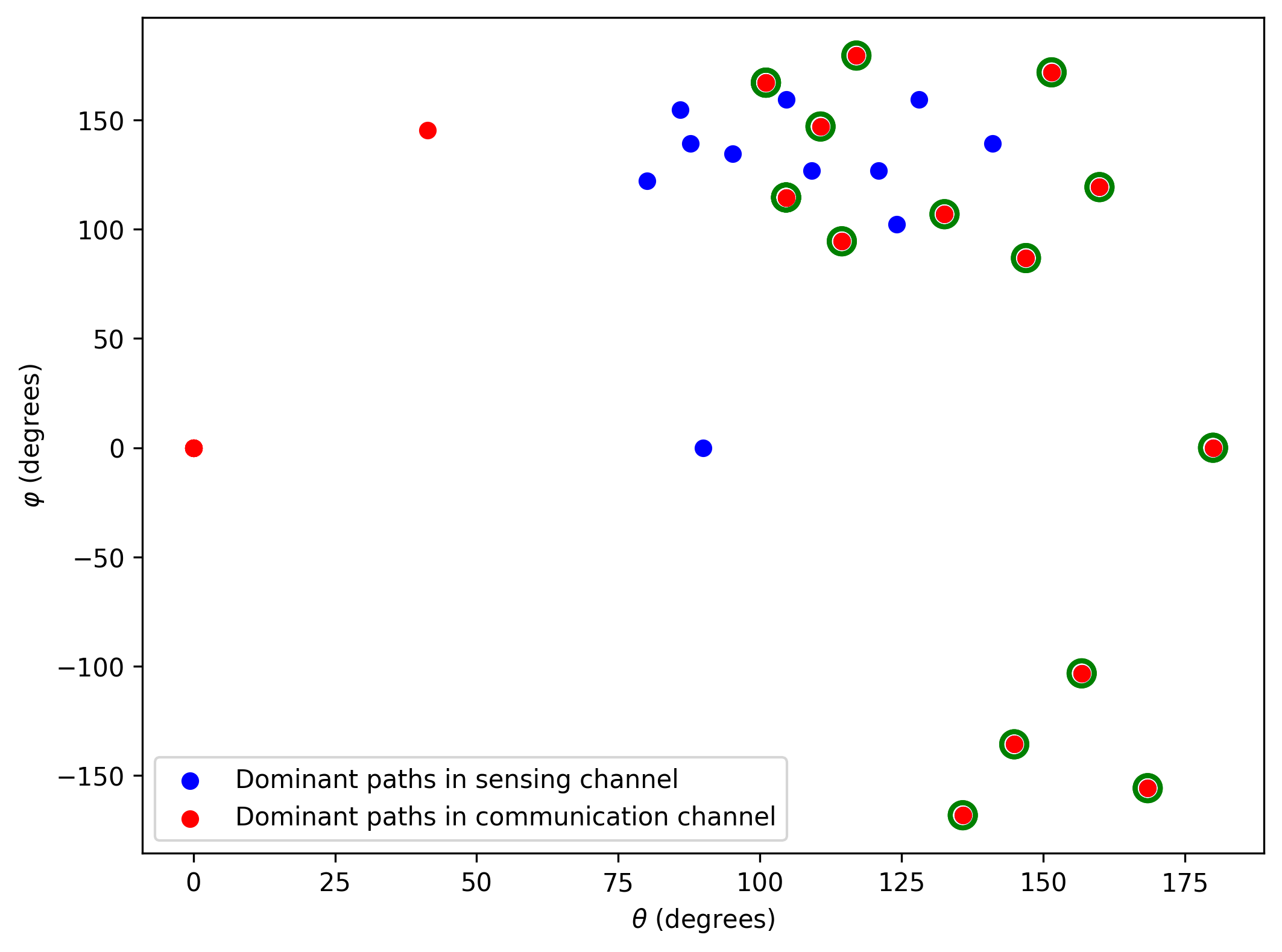}
  \caption*{(b) Path angle distribution.}
\end{minipage}
\caption{Angle distribution of dominant signal paths for communication and sensing channels in a street environment with moving vehicles.}
\label{Fig.Angle}
\end{figure}

By comparing the communication channel $\boldsymbol{h}_{\mathsf{c},t} (\tau )$ in (\ref{communication channel}) and the sensing channel $\boldsymbol{H}_{\mathsf{s},t} (\tau )$ in (\ref{sensing channel}), it is observed that they have the same parameters $\{f_{\mathsf{d},i}, \theta _{i},\varphi _{i}\}$ for the paths associated with the shared scatterers in $\mathcal{N}_\mathsf{sha}$, which lead to overlapping signal paths with the same elevation and azimuth angles. To verify this phenomenon, we conduct ray-tracing simulations using the NVIDIA Sionna platform \cite{sionna} by considering the street environment with one ISAC BS and one moving vehicle as MU, in which buildings, grounds, and other mobile vehicles collectively generate channel paths as shown in Fig. \ref{Fig.Angle}(a). Fig. \ref{Fig.Angle}(b) shows the distribution of azimuth and elevation angles of the dominant signal paths for communication and sensing channels. It is observed that there are 16 and 26 dominant paths for communication and sensing channels, respectively. Among them, 14 paths are overlapping with the same azimuth and elevation angles, which correspond to shared scatterers. This observation implies that the sensing channel $\boldsymbol{H}_{\mathsf{s},t} (\tau )$ actually contains useful information about the communication channel $\boldsymbol{h}_{\mathsf{c},t} (\tau )$. This motivates us to improve channel prediction accuracy for communication CSI by exploiting the correlation with sensing channels. 






Our objective is to predict future communication CSI over the upcoming $Q$ slots, based on historical observation of both communication and sensing CSI over the previous $P$ slots. For notational convenience, we denote the frequency-domain communication and sensing channels at each slot $i$ as $\tilde{\boldsymbol{h}}_{\mathsf{c},i} = \left[ \bar{\boldsymbol{h}}_{\mathsf{c},i}^T[1], \ldots, \bar{\boldsymbol{h}}_{\mathsf{c},i}^T[K] \right]^{T}$ and $\tilde{\boldsymbol{H}}_{\mathsf{s},i}=\left [ \bar{\boldsymbol{H}}_{\mathsf{s},i}^T[1],\ldots , \bar{\boldsymbol{H}}^T_{\mathsf{s},i}[K] \right ]^{T}$, respectively. Therefore, at each slot $t$, the sensing-assisted channel prediction problem corresponds to finding a mapping function \( \mathcal{F}(\cdot) \) to predict the communication CSI $\{\tilde{\boldsymbol{h}}_{\mathsf{c},i}\}_{i=t+1}^{t+Q}$ in the upcoming $Q$ slots based on the communication CSI $\{\tilde{\boldsymbol{h}}_{\mathsf{c},i}\}_{i=t-P+1}^{t}$ and sensing CSI $\{\tilde{\boldsymbol{H}}_{\mathsf{s},i}\}_{i=t-P+1}^{t}$, i.e., 
\begin{align}\label{channel prediction}
    \{\tilde{\boldsymbol{h}}_{\mathsf{c},i}\}_{i=t+1}^{t+Q} = \mathcal{F} \big( 
    \{\tilde{\boldsymbol{h}}_{\mathsf{c},i}\}_{i=t-P+1}^{t}, \{\tilde{\boldsymbol{H}}_{\mathsf{s},i}\}_{i=t-P+1}^{t}\big).
\end{align}
\footnote{It is assumed that the ISAC BS perfectly knows historical CSI $\{\tilde{\boldsymbol{h}}_{\mathsf{c},i}\}_{i=t-P+1}^{t}$ and $ \{\tilde{\boldsymbol{H}}_{\mathsf{s},i}\}_{i=t-P+1}^{t}$ based on, e.g., channel estimation. This assumption is made to focus our study on the channel prediction. The effect of imperfect CSI will be analyzed in future work.} However, due to the inherent complexity and dynamic spatiotemporal variations of wireless environments, it is a challenging task to find a good nonlinear mapping function \( \mathcal{F}(\cdot) \) in problem (\ref{channel prediction}), and conventional model-based designs are generally infeasible, especially when the number of scatterers or signal paths becomes sufficiently large in rich scattering environments.
\section{LLM-based Sensing-assisted Channel Prediction Approach}
\begin{figure*}[!t]
\setlength{\abovecaptionskip}{-0pt}
\setlength{\belowcaptionskip}{-20pt}
\centering
\includegraphics[width=0.9\textwidth]{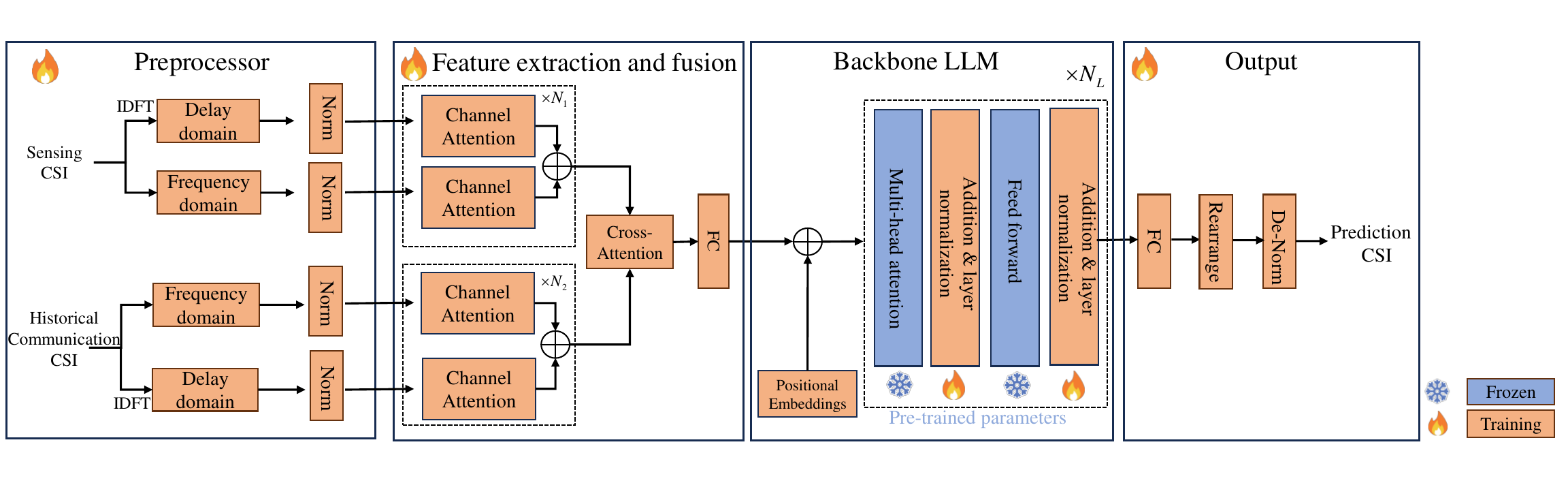}
\caption{The network architecture of the proposed approach.}
\label{Fig.Net}
\end{figure*}

In this section, we propose an LLM-based sensing-assisted channel prediction approach to address the above challenge, in which a pre-trained text-based LLM is  adapted to jointly utilize the historical communication and sensing CSI data in complex matrix format for predicting future communication CSI. Towards this end, we design several specific modules, including the preprocessor, feature extraction and fusion, backbone, and output modules. The overall network structure of our design is shown in Fig. \ref{Fig.Net}, for which the details are discussed in the following. Note that adapting LLM for channel prediction (LLM4CP) has been studied in \cite{liu2024llm4cp} without the sensing assistance. Different from the design in \cite{liu2024llm4cp}, our proposed approach uses the convolutional long short term memory (ConvLSTM) \cite{convlstm} as the channel attention module to extract features from the CSI data and the cross-attention mechanism to fuse features from sensing and communication CSI, thus achieving improved performance.
\vspace{-10pt}
\subsection{Preprocessor Module}
The preprocessor module transforms the frequency-domain CSI into the delay-domain representation and performs the CSI data normalization for both of them to facilitate the processing. In particular, the frequency-domain CSI $\tilde{\boldsymbol{h}}_{\mathsf{c},t}$ and $\tilde{\boldsymbol{H}}_{\mathsf{s},t}$ primarily describe the signal amplitude and phase characteristics across subcarriers and time slots. To better leverage the relevant information contained, we first apply the $K$-point inverse DFT (IDFT) to convert communication CSI $\{\bar{\boldsymbol{h}}_{\mathsf{c},i}[k]\}_{k=1}^K$ and sensing CSI $\{\bar{\boldsymbol{H}}_{\mathsf{s},i}[k]\}_{k=1}^K$ from the frequency domain into the delay-domain representation $\{\hat{\boldsymbol{h}}_{\mathsf{c},i}[\hat{\tau}]\}_{\hat{\tau}=1}^K$ and $\{\hat{\boldsymbol{H}}_{\mathsf{s},i}[\hat{\tau}]\}_{\hat{\tau}=1}^K$, respectively. Furthermore, to reduce the computational complexity for training, we implement the parallel processing to handle the CSI for each pair of transmit and receive antennas separately. In each parallel processing for antenna $n \in \{1,\ldots, N\}$, we use the $n$-th elements of delay-domain $\hat{\boldsymbol{h}}_{\mathsf{c},i}[\hat{\tau}]$ and frequency-domain $\bar{\boldsymbol{h}}_{\mathsf{c},i}[k]$, as well as the $n$-th diagonal elements of delay-domain $\hat{\boldsymbol{H}}_{\mathsf{s},i}[\hat{\tau}]$ and frequency-domain $\{\bar{\boldsymbol{H}}_{\mathsf{s},i}[k]\}$ over slots $i\in\{t-P+1,\ldots,t\}$ and $\tau\in\{1,\ldots,K\}$ as the network input, forming complex matrices $\boldsymbol{X}_{\mathsf{c},n}, \boldsymbol{X}_{\mathsf{c},n,\tau}, \boldsymbol{X}_{\mathsf{s},n}, \boldsymbol{X}_{\mathsf{s},n,\tau} \in \mathbb{C}^{K \times P}$.

In addition, as neural networks generally deal with real numbers, we convert all complex-valued representations $\boldsymbol{X}_{\mathsf{c},n}$, $\boldsymbol{X}_{\mathsf{c},n,\tau}$, $\boldsymbol{X}_{\mathsf{s},n}$, and $\boldsymbol{X}_{\mathsf{s},n,\tau}$ into real tensors and normalize them as $\bar{\boldsymbol{X}}_{\mathsf{c},n,\tau}$, $\bar{\boldsymbol{X}}_{\mathsf{c},n}$, $\bar{\boldsymbol{X}}_{\mathsf{s},n,\tau}$, and $\bar{\boldsymbol{X}}_{\mathsf{s},n}\in \mathbb{R}^{2 \times K \times P}$, respectively, to facilitate the training.
\vspace{-10pt}
\subsection{Feature Extraction and Fusion Module}
The feature extraction and fusion module employs channel attention and cross-attention to capture and fuse the spatiotemporal features of communication and sensing CSI, in which $\bar{\boldsymbol{X}}_{\mathsf{c},n,\tau}$, $\bar{\boldsymbol{X}}_{\mathsf{c},n}$, $\bar{\boldsymbol{X}}_{\mathsf{s},n,\tau}$ 
,~and $\bar{\boldsymbol{X}}_{\mathsf{s},n}$ are taken as the input. First, the channel attention module consists of several ConvLSTM cells, each of which integrates convolutional neural networks (CNN) with long short-term memory (LSTM) units. As compared to the conventioanl LSTM gates, the ConvLSTM design can better capture both temporal and spatial correlations in sensing and communication CSI data. Note that the channel attention modules are concatenated multiple times to enhance the effectiveness of feature extraction. Next, we use the cross-attention mechanism to fuse the sensing and communication features instead of simply adding them. The output after the channel attention and cross-attention processing is given by 
\begin{equation}
\boldsymbol{X}_{\text{CA},n} = \text{CA} (\text{CL}^{(N_1)}(\bar{\boldsymbol{X}}_{\mathsf{s},n,\tau},\bar{\boldsymbol{X}}_{\mathsf{s},n}) + \text{CL}^{(N_2)}(\bar{\boldsymbol{X}}_{\mathsf{c},n,\tau},\bar{\boldsymbol{X}}_{\mathsf{c},n})),
\end{equation}
with $\boldsymbol{X}_{\text{CA},n} \in \mathbb{R}^{2K \times P}$, where $\text{CL}^{(N)}(\cdot)$ denotes the channel attention cascaded $N$ times, and $\text{CA}(\cdot)$ represents the cross-attention. 
Furthermore, we employ a single fully connected (FC) layer to map $\boldsymbol{X}_{\text{CA},n}$ into $\bar{\boldsymbol{X}}_{\text{CA},n} \in \mathbb{R}^{F \times P}$, where $F$ is the feature dimension of the pre-trained LLM.


\subsection{Backbone Network and Output}
After feature extraction and fusion, we perform non-learnable positional encoding $\boldsymbol{X}_{\text{PE},n} \in \mathbb{R}^{F \times P}$ to preserve the temporal and spatial information in the input sequences. This ensures that the model can account for the sequential and spatial relationships within the CSI data, thereby enhancing its ability to learn spatiotemporal patterns. Therefore, the embedding $\boldsymbol{X}_{\text{EB},n} \in \mathbb{R}^{F \times P}$ is obtained as
\setlength{\abovedisplayskip}{1pt}
\setlength{\belowcaptionskip}{1pt}
\begin{equation}
\boldsymbol{X}_{\text{EB},n} = \bar{\boldsymbol{X}}_{\text{CA},n} + \boldsymbol{X}_{\text{PE},n} .
\end{equation}
Next, we feed  the embeddings of CSI data $\boldsymbol{X}_{\text{EB},n}$ into the LLM backbone network as preprocessed CSI ``tokens''. While different LLMs (such as Llama \cite{touvron2023llama}) are implementable here, we use GPT-2\cite{gpt2} as our backbone.  
The GPT-2 backbone consists of a learnable positional embedding layer and stacked transformer decoders, each layer of which consists of multi-head attention layers, feed forward layers, addition, and layer normalization, as shown in Fig. \ref{Fig.Net}. The output of the LLM backbone is expressed as 
\begin{equation}
\boldsymbol{X}_{\text{LLM},n} = \text{LLM}(\boldsymbol{X}_{\text{EB},n}).
\end{equation}
After the LLM processing, we design an output module to convert the LLM's output features $\boldsymbol{X}_{\text{LLM},n}\in \mathbb{R}^{F \times Q}$ into the final prediction results. In particular, we first use an FC layer to transform the dimensions of the LLM's output layer, and then rearrange them and perform de-normalization to generate the final output of the network $\boldsymbol{{X}}_{\text{out},n} \in \mathbb{R}^{2 \times K \times Q}$, i.e., 
\setlength{\abovedisplayskip}{1pt}
\setlength{\belowcaptionskip}{1pt}
\begin{equation}
{\boldsymbol{{X}}}_{\text{out},n} =\text{De-Norm}( \text{FC}(\boldsymbol{X}_{\text{LLM},n})).
\end{equation}
The final prediction result $\hat{\boldsymbol{H}} _{\mathsf{c},n}\in \mathbb{C}^{K \times Q}$ is then obtained by $\hat{\boldsymbol{H}}_{\mathsf{c},n} = \boldsymbol{X}_{\text{out},n}[1, :, :] + \mathrm{j} \boldsymbol{X}_{\text{out},n}[2, :, :]$.
\subsection{Training Configuration}
Finally, we discuss the training of the considered LLM-based network. 
In the training phase, the normalized mean square error (NMSE) is adopted as the loss function to minimize the prediction error, i.e.,
\begin{equation}
\mathcal{L}_\text{NMSE} = \frac{\sum_{n=1}^{N} \|\hat{\boldsymbol{H}}_{\mathsf{c},n} - \boldsymbol{H}_{\mathsf{c},n}\|_F^2}{\sum_{n=1}^{N} \|\boldsymbol{H}_{\mathsf{c},n}\|_F^2},
\end{equation} 
where $\hat{\boldsymbol{H}}_{\mathsf{c},n}$ and $\boldsymbol{H}_{\mathsf{c},n} \in \mathbb{C}^{K \times Q}$ correspond to the predicted and ground-truth CSI matrices for the $n$-th antenna pair, $N$ is the total number of antenna pairs, respectively, and $\|\cdot\|_F$ denotes the Frobenius norm. In addition, the validation loss also adopts the same loss function. It is worth noting that during training, to preserve the general knowledge of the pre-trained model, the multi-head attention layer and the feedforward neural network layer are frozen, while the addition, layer normalization, and positional embedding layers are fine-tuned to adapt the LLM to the channel prediction task.
\vspace{-0.5em}

\section{Experiment Results}
This section presents experiment results to validate the performance of our proposed LLM-based sensing-assisted channel prediction design. First, we explain the generation process of the simulation dataset. The simulated CSI dataset is generated using the Sionna platform's ray-tracing for a specified environmental geometry, modeling multi-path characteristics based on material parameters \cite{sionna}. We consider a multi-antenna OFDM system, where the BS is equipped with a dual-polarized and half-wavelength UPA with $N_{v} = N_{h} =4$, and the MU is equipped with a single omnidirectional antenna. The system is configured with a 28 GHz carrier frequency and 5.76 MHz bandwidth, with 48 subcarriers at 60 kHz spacing. We aim to predict the communication CSI in the future $Q =5$ slots based on historical CSI in the previous $P=10$ slots. The MU has a velocity uniformly distributed between 10km/h and 100 km/h. The dataset contains 6000 training samples and 600 test samples. 

More specifically, we consider a street scenario with multiple moving vehicles as scatterers. The ISAC BS is placed at a height of 10 m. We consider three wireless signal propagation mechanisms to simulate real-world channels, including LoS, reflection, and scattering. The initial ray directions are uniformly distributed on the unit sphere based on a Fibonacci lattice, ensuring broad coverage and uniform distribution of the rays. By integrating a shoot-and-bounce strategy, the method efficiently identifies the intersection points between rays and objects within the scene, precisely capturing complex multi-path effects. Since the movement of objects during the considered time slots is relatively small, we consider the AoD and path delays to be invariant. Based on the speed of vehicles in the scenario and the speed of the MU, Doppler frequency shifts are applied to all paths to simulate the time variation of sensing and communication channels.

To validate the performance of the proposed approach, we consider the scheme without sensing assistance, as well as several commonly adopted deep learning-based channel prediction methods as baselines.
\begin{itemize}
\item [$\bullet$] No sensing assistance: This scheme only uses historical communication CSI for channel prediction without using any sensing information.
\item[$\bullet$] LSTM: LSTM is designed with memory cells and multiplicative gates to deal with long-term dependency. 
\item[$\bullet$] Transformer \cite{transformer}: A transformer-based parallel predictor is utilized for channel prediction.
\item[$\bullet$] CNN: A CNN-based channel predictor is considered, which treats the prediction process of time-frequency CSI data as a 2D image processing task.
\end{itemize} 
In order to ensure fairness, all the aforementioned deep learning-based methods process antenna dimensions in parallel and adopt the NMSE as the loss function for training.
\begin{figure}
\setlength{\abovecaptionskip}{-0pt}
\setlength{\belowcaptionskip}{-10pt}
\centering
\includegraphics[width=6.4cm]{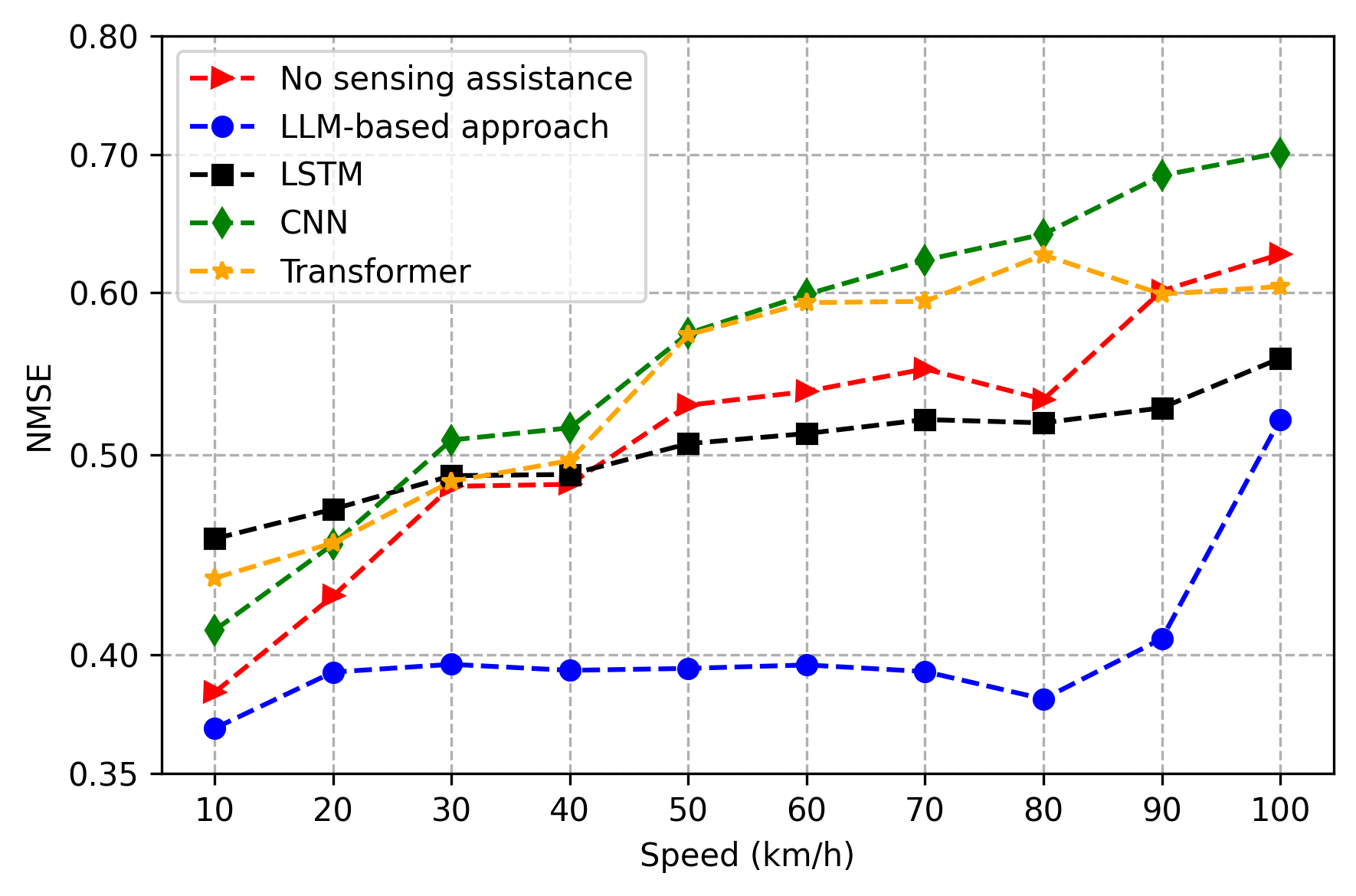}
\caption{The NMSE performance versus MU speed.}
\label{Fig.speed}
\end{figure}
\begin{figure}
\setlength{\abovecaptionskip}{-0pt}
\setlength{\belowcaptionskip}{-15pt}
\centering
\includegraphics[width=6.4cm]{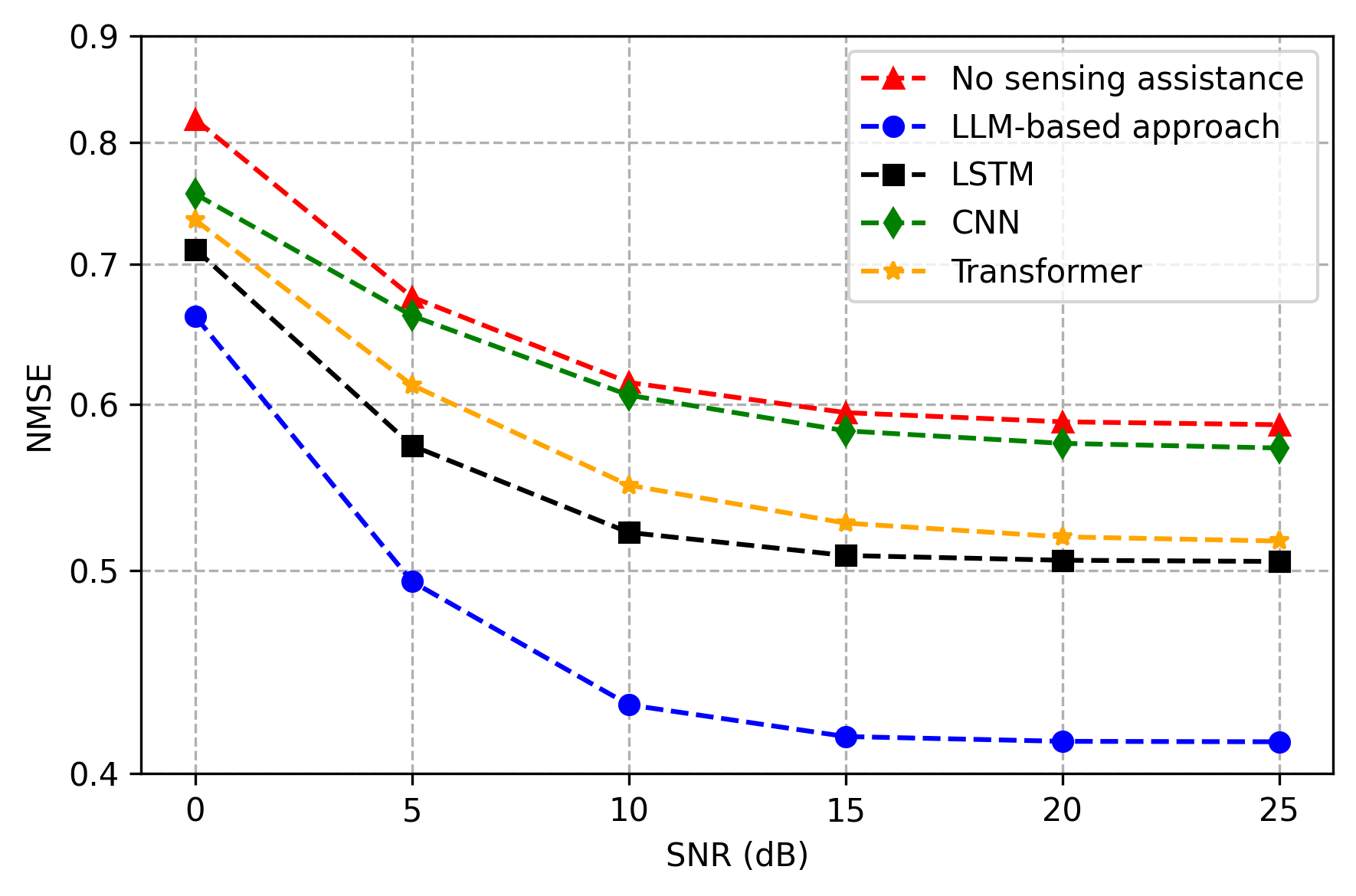}
\caption{The NMSE performance versus SNR with noisy historical CSI.}
\label{Fig.snr}
\end{figure}

Fig. \ref{Fig.speed} shows the NMSE performance versus the speed of MU. It is observed that our proposed LLM-based approach consistently surpasses other benchmark schemes by achieving the lowest NMSE throughout the whole speed range, demonstrating its robustness in dynamic channel conditions. The proposed approach optimizes communication channel prediction by leveraging environmental sensing data, proving particularly advantageous in high-mobility scenarios with rapidly changing channel characteristics. In contrast, the scheme without sensing assistance, exhibits inferior performance across all velocities, underscoring the critical role of sensing information in maintaining prediction accuracy in practical complex environments.

Fig. \ref{Fig.snr} shows the NMSE performance versus the signal-to-noise ratio (SNR) with noisy historical CSI. Specifically, in the testing phase, the historical CSI data is added by white Gaussian noise with variance $\sigma_n^2$ and the SNR is defined as $1/\sigma_n^2$. To enhance the robustness, during the training phase, the SNR is set to be uniformly distributed between 0 and 25 dB for all baselines. It is observed that for all schemes, lower SNR results in higher prediction NMSE. Notably, the proposed approach exhibits the lowest NMSE at the whole SNR regime. This demonstrates that integrating sensing data with the LLM framework effectively captures the dynamic characteristics of the wireless channel, resulting in high robustness against CSI noise.

In addition, to validate the effectiveness of several specific modules, we conduct ablation experiments by removing relevant modules. Table \ref{tab:ablation experiments} shows the NMSE performance over all testing velocities. It is observed that the removal of any of these four modules results in a loss of performance, indicating the necessity of these modules for high predictive accuracy.
\begin{table}
\centering
\caption{Ablation Experiment Results.}
\label{tab:ablation experiments}
\resizebox{\columnwidth}{!}{
\begin{tabular}{c|c|c|c|c|c}
\hline
\textbf{Metric} & Our approach & W/o sensing & W/o channel attention & W/o cross attention & W/o LLM \\ \hline
NMSE & \textbf{0.407} & 0.515 & 0.504 & 0.446 & 0.469 \\ \hline
\end{tabular}
}
\end{table}
\vspace{-10pt}
\section{Conclusion}
This letter proposed a novel LLM-based sensing-assisted channel prediction framework for multi-antenna OFDM systems operating in complex wireless environments. By exploiting the inherent environmental correlation between mono-static sensing and bi-static communication channels, this approach predicts future communication CSI by jointly learning spatiotemporal patterns from historical sensing and communication CSI data. By adapting pre-trained text-based LLMs to handle complex-matrix-form CSI data, the approach capitalizes on the LLM's superior capability to model intricate relationships across time, frequency, and spatial domains, and achieves significant performance gains over both non-sensing-assisted benchmarks and existing deep learning solutions.
\vspace{-10pt}
\bibliographystyle{ieeetr}
\bibliography{ref}

\begin{thebibliography}{10}

\bibitem{cui.shuguang_large-scale_mimo}
Z.~Wang, J.~Zhang, H.~Du, D.~Niyato, S.~Cui, B.~Ai, M.~Debbah, K.~B. Letaief, and H.~V. Poor, ``A tutorial on extremely large-scale {MIMO} for {6G}: Fundamentals, signal processing, and applications,'' {\em IEEE Commun. Surv. Tutor.}, vol.~26, no.~3, pp.~1560--1605, Jan. 2024.

\bibitem{4.yin2020addressing}
H.~Yin, H.~Wang, Y.~Liu, and D.~Gesbert, ``Addressing the curse of mobility in massive {MIMO} with prony-based angular-delay domain channel predictions,'' {\em IEEE J. Sel. Areas Commun.}, vol.~38, no.~12, pp.~2903--2917, Dec. 2020.

\bibitem{liu2024llm4cp}
B.~Liu, X.~Liu, S.~Gao, X.~Cheng, and L.~Yang, ``{LLM4CP}: Adapting large language models for channel prediction,'' {\em J. Commun. Inf. Networks}, vol.~9, no.~2, pp.~113--125, Jun. 2024.

\bibitem{transformer}
H.~Jiang, M.~Cui, D.~W.~K. Ng, and L.~Dai, ``Accurate channel prediction based on transformer: {Making} mobility negligible,'' {\em IEEE J. Sel. Areas Commun.}, vol.~40, no.~9, pp.~2717--2732, Sep. 2022.

\bibitem{convlstm}
G.~Liu, Z.~Hu, L.~Wang, J.~Xue, H.~Yin, and D.~Gesbert, ``Spatio-temporal neural network for channel prediction in massive {MIMO-OFDM} systems,'' {\em IEEE Trans. Commun.}, vol.~70, no.~12, pp.~8003--8016, Dec. 2022.

\bibitem{7.liu2022integrated}
F.~Liu, Y.~Cui, C.~Masouros, J.~Xu, T.~X. Han, Y.~C. Eldar, and S.~Buzzi, ``Integrated sensing and communications: Toward dual-functional wireless networks for {6G} and beyond,'' {\em IEEE J. Sel. Areas Commun.}, vol.~40, no.~6, pp.~1728--1767, Jun. 2022.

\bibitem{ren2024sensing}
Z.~Ren, L.~Qiu, J.~Xu, and D.~W.~K. Ng, ``Sensing-assisted sparse channel recovery for massive antenna systems,'' {\em IEEE Trans. Veh. Technol.}, vol.~73, no.~11, pp.~17824--17829, Nov. 2024.

\bibitem{8_liu_ywj}
F.~Liu, W.~Yuan, C.~Masouros, and J.~Yuan, ``Radar-assisted predictive beamforming for vehicular links: Communication served by sensing,'' {\em IEEE Trans. Wireless Commun.}, vol.~19, no.~11, pp.~7704--7719, Aug. 2020.

\bibitem{sensing-assisted_bf}
Y.~Zhao, X.~Xu, Y.~Zeng, F.~Liu, Y.~Huang, and Y.~L. Guan, ``Sensing-assisted predictive beamforming with multipath echo signals,'' {\em IEEE Trans. Veh. Technol.}, pp.~1--15, early access, Feb. 04, 2025.
\newblock doi:{10.1109/TVT.2025.3530641}.

\bibitem{15.su2024large}
J.~Su, C.~Jiang, X.~Jin, Y.~Qiao, T.~Xiao, H.~Ma, R.~Wei, Z.~Jing, J.~Xu, and J.~Lin, ``Large language models for forecasting and anomaly detection: A systematic literature review,'' {\em arXiv preprint arXiv:2402.10350}, 2024.

\bibitem{timellm}
M.~Jin, S.~Wang, L.~Ma, Z.~Chu, J.~Y. Zhang, X.~Shi, P.-Y. Chen, Y.~Liang, Y.-F. Li, S.~Pan, and Q.~Wen, ``{Time-LLM}: Time series forecasting by reprogramming large language models,'' in {\em Proc. Int. Conf. Learn. Represent.}, 2024.

\bibitem{goldsmith2005wireless}
A.~Goldsmith, {\em Wireless Communications}.
\newblock Cambridge, U.K.: Cambridge Univ. Press, 2005.

\bibitem{sionna}
J.~Hoydis, S.~Cammerer, F.~{Ait Aoudia}, A.~Vem, N.~Binder, G.~Marcus, and A.~Keller, ``Sionna: An open-source library for next-generation physical layer research,'' {\em arXiv preprint arXiv:2203.11854}, 2022.

\bibitem{touvron2023llama}
H.~Touvron, L.~Martin, K.~Stone, P.~Albert, A.~Almahairi, Y.~Babaei, N.~Bashlykov, S.~Batra, P.~Bhargava, S.~Bhosale, {\em et~al.}, ``Llama 2: Open foundation and fine-tuned chat models,'' {\em arXiv preprint arXiv:2307.09288}, 2023.

\bibitem{gpt2}
A.~Radford, J.~Wu, R.~Child, D.~Luan, D.~Amodei, I.~Sutskever, {\em et~al.}, ``Language models are unsupervised multitask learners,'' {\em OpenAI blog}, vol.~1, no.~8, p.~9, 2019.

\end{thebibliography}

\end{document}